\documentstyle[twoside,fleqn,espcrc2,epsf]{article}

\newcommand{\AmS}{{\protect\the\textfont2
  A\kern-.1667em\lower.5ex\hbox{M}\kern-.125emS}}

\newcommand{\be}{\begin{equation}}
\newcommand{\ee}{\end{equation}}
\newcommand{\ben}{\begin{eqnarray}}
\newcommand{\een}{\end{eqnarray}}

\title{Two-Flavor Chiral Phase Transition in Lattice QCD
with the Kogut-Susskind Quark Action\thanks{poster presented by A. Ukawa}}

\author{JLQCD Collaboration :
	S. Aoki\address{Institute of Physics, University of Tsukuba,
        Tsukuba, Ibaraki 305, Japan},
        M. Fukugita\address{Institute for Cosmic Ray Research,
        University of Tokyo, Tanashi, Tokyo 188, Japan},
        S. Hashimoto\address{Computing Research Center,
        High Energy Accelerator Research Organization (KEK),\\
        Tsukuba, Ibaraki 305, Japan},
        N. Ishizuka$^{\rm a,d}$,
	Y. Iwasaki$^{\rm a,d}$,\\
	K. Kanaya$^{\rm a,}$\address{Center for Computational Physics,
        University of Tsukuba, Tsukuba, Ibaraki 305, Japan},
	Y. Kuramashi\address{Institute of Particle and Nuclear Studies,
        High Energy Accelerator Research Organization (KEK),
        Tsukuba, Ibaraki 305, Japan},
        H. Mino\address{Faculty of Engineering, Yamanashi University,
        Kofu 400, Japan},
	M. Okawa$^{\rm e}$,
	A. Ukawa$^{\rm a}$,
	T. Yoshi\'e$^{\rm a,d}$
}

\begin{document}

\begin{abstract}

A summary is presented of a scaling study of
the finite-temperature chiral phase
transition of two-flavor QCD with the Kogut-Susskind quark action
based on simulations on $L^3\times4$ ($L$=8, 12 and 16) lattices at
the quark mass of $m_q=0.075, 0.0375, 0.02$ and 0.01.
We find a phase transition to be absent for
$m_q\geq 0.02$, and also quite likely at $m_q=0.01$.
The quark mass dependence of susceptibilities
is consistent with a second-order transition at $m_q=0$.
The exponents, however, deviate from
the O(2) and O(4) values theoretically expected.

\end{abstract}

\maketitle

\section{Introduction}

The order of the two-flavor chiral phase transition is a
basic question in finite-temperature lattice QCD.
Earlier finite-size studies disfavored
a first-order transition, indicating the chiral transition being
second-order at a zero quark mass\cite{founf2,columbianf2}.
This was also corroborated by a
scaling analysis of Ref.~\cite{karschlaermann}.
To advance the scaling argument, we have pushed forward
simulations toward larger spatial lattice sizes and
smaller quark masses, where we found\cite{preliminary}
exponents differing from the earlier results\cite{karschlaermann}.
We have since completed our runs\cite{fullpaper}, and here present
a summary of results.  For recent similar attempts we refer to
Refs.~\cite{laermann,toussaint}.

\section{Simulation}

Our study is made with the plaquette gauge action and the
Kogut-Susskind quark action on lattices of a size $L^3\times 4$ with
$L=8, 12, 16$ at the quark masses $m_q=0.075, 0.0375, 0.02, 0.01$.
For each set $(L, m_q)$, 10000 trajectories of unit length are generated
by the hybrid R algorithm for a single value of $\beta$ close to the
transition.  The standard reweighting technique is employed
to calculate the $\beta$ dependence of observables around the
simulation point.
The following susceptibilities are calculated at each trajectory:
\ben
\chi_m&=&V\left[\langle\left(\overline{\psi}\psi\right)^2\rangle-
\langle\overline{\psi}\psi\rangle^2\right],\\
\chi_{t,f}&=&V\left[\langle \left(\overline{\psi}\psi\right)
                            \left(\overline{\psi}D_0\psi\right)\rangle
             -\langle\overline{\psi}\psi\rangle
                            \langle\overline{\psi}D_0\psi\rangle\right]
                \label{eq:sustebegin}\\
\chi_{t,i}&=&V\left[\langle \left(\overline{\psi}\psi\right)P_i\rangle
             -\langle\overline{\psi}\psi\rangle
                            \langle P_i\rangle\right],\\
\chi_{e,f}&=&V\left[\langle\left(\overline{\psi}D_0\psi\right)^2\rangle-
\langle\overline{\psi}D_0\psi\rangle^2\right],\\
\chi_{e,i}&=&V\left[\langle \left(\overline{\psi}D_0\psi\right)P_i\rangle
             -\langle\overline{\psi}D_0\psi\rangle
                            \langle P_i\rangle\right],\\
\chi_{e,ij}&=&V\left[\langle P_iP_j\rangle
            -\langle P_i\rangle\langle P_j\rangle\right],
\label{eq:susteend}
\een
where $V=L^3\cdot 4$ denotes the lattice volume,
$D_0$ the temporal component of the Kogut-Susskind
operator, $i,j=\sigma, \tau$, and $P_{\sigma,\tau}$ the spatial and
temporal plaquette.  Averages are taken over the last 8000 trajectories,
with errors estimated through jackknife analyses with the bin size of
800 trajectories.

\section{Finite-size scaling analysis}

\begin{figure}[bt]
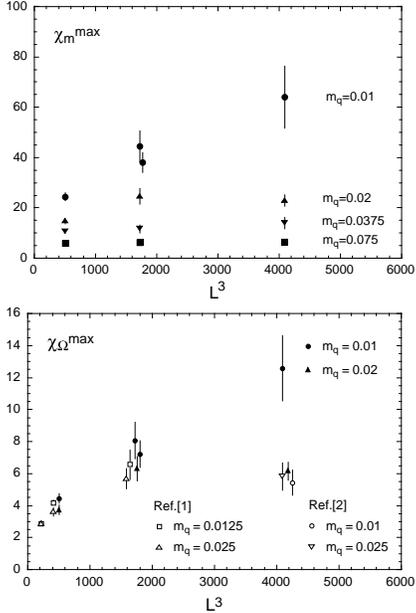

\centerline{\epsfxsize=55mm \epsfbox{fig1.epsf}}
%\vspace*{-8mm}
\centerline{\epsfxsize=55mm \epsfbox{fig1b.epsf}}
\vspace*{-8mm}
\caption{Peak height of $\chi_m$ and $\chi_\Omega$
as a function of spatial volume $L^3$.  Two runs are made at
$L=12$ and $m_q=0.01$.  For $\chi_\Omega$ data plotted with open symbols are
results from previous studies\cite{founf2,columbianf2}.}
\label{fig:heightvsvolume}
\vspace*{-8mm}
\end{figure}

\begin{figure}[bt]
\centerline{\epsfxsize=70mm \epsfbox{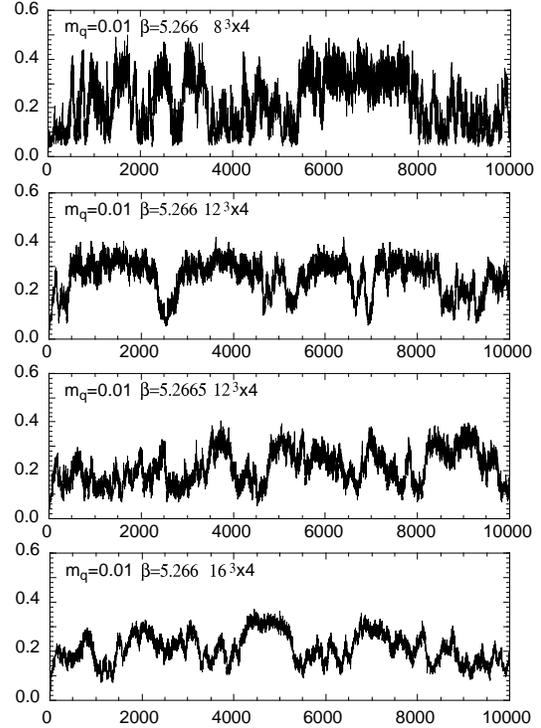}}
\vspace*{-8mm}
\caption{Time history of $\overline{\psi}\psi$ for the runs with $m_q$=0.01.
Two runs made for $L=12$ are both shown.}
\label{fig:histories}
\vspace*{-8mm}
\end{figure}

In Fig.~\ref{fig:heightvsvolume} we plot the results for the spatial
volume dependence of the peak height of the susceptibilities $\chi_m$
and $\chi_\Omega$.  For
the latter, results from previous studies are also shown for comparison
(open symbols).
For $m_q\geq 0.02$ the two susceptibilities do not increase beyond the
size $L=12$, supporting the previous conclusion of an absence of phase
transition down to $m_q=0.025$\cite{founf2,columbianf2}.

For $m_q=0.01$ an increase of the peak height continues up to $L=16$ in
a linear manner.
This result is quite different from that of an earlier work\cite{columbianf2}
which showed a flattening behavior for $L=16$ (see open circle for
$\chi_\Omega^{max}$).
We may ascribe this underestimate to a smaller statistics
(2500 trajectories\cite{columbianf2}
as compared to 10000 employed here).

A linear increase of susceptibilities is a behavior
expected for a first-order transition.  However, examining the
time histories of chiral order parameter for $m_q=0.01$
shown in Fig.~\ref{fig:histories},
we observe that metastability signals become weaker toward larger
spatial sizes:  a flip-flop behavior is most apparent
for $L=8$, while irregular fluctuations are more dominant for larger $L$.
Correspondingly, the histogram of
$\overline{\psi}\psi$ clearly shows a double-peak distribution for $L=8$,
which is less evident for $L=12$ and hardly visible for $L=16$.
Furthermore, if we normalizes $L$ by the pion correlation length
$\xi_\pi=1/m_\pi$ at zero temperature, we find the increase of
susceptibilities at $m_q=0.01$ for $L=12-16$ being similar to that
at $m_q=0.02$ for $L=8-12$.

We conclude that the increase of susceptibilities for $m_q=0.01$ is
probably due to insufficient spatial volume, and that a phase transition
is also likely to be absent for $m_q=0.01$.

\section{Second-order scaling analysis}

\begin{table}[t]
\begin{center}
\setlength{\tabcolsep}{0.2pc}
\caption{Critical exponents for each spatial size $L$ as compared to
O(2), O(4) and mean-field (MF) values. Critical coupling obtained
from $\chi_m$ is employed to extract $z_g$.}
\label{tab:exponents}
\vspace*{2mm}
\begin{tabular}{lllllll}
\hline
	&O(2)	&O(4)	&MF			&$L=8$	&$L=12$	&$L=16$\\
\hline
$z_g$	&0.60 	&0.54	&2/3			&0.70(11)&0.74(6)&0.64(5)\\
\hline
$z_m$	&0.79	&0.79	&2/3			&0.70(4)&0.99(8)&1.03(9)\\
\hline
$z_t$	&0.39	&0.33	&1/3		\\
$z_{t,f}$		&	&	&	&0.42(5)&0.75(9)&0.78(10)\\
$z_{t,\sigma}$		&	&	&	&0.47(5)&0.81(10) &0.82(12)\\
$z_{t,\tau}$		&	&	&	&0.47(5)&0.81(9) &0.83(12)\\
\hline
$z_e$	&-0.01	&-0.13	&0		\\
$z_{e,f}$		&	&	&	&0.21(4)&0.28(7)&0.38(7)\\
$z_{e,\sigma}$		&	&	&	&0.25(6)&0.56(11) &0.58(13)\\
$z_{e,\tau}$		&	&	&	&0.22(6)&0.52(10) &0.55(12)\\
$z_{e,\sigma\sigma}$	&	&	&	&0.18(5)&0.46(8) &0.43(10)\\
$z_{e,\sigma\tau}$	&	&	&	&0.20(5)&0.51(9) &0.50(12)\\
$z_{e,\tau\tau}$	&	&	&	&0.19(5)&0.48(9) &0.47(11)\\
\hline
\end{tabular}
\end{center}
\vspace*{-5mm}
\end{table}

For a second-order transition, the critical coupling and susceptibilities
are expected to exhibit a scaling behavior toward $m_q\to 0$ given by
\ben
g_c^{-2}(m_q)&=&g_c^{-2}(0)+c_gm_q^{z_g} \label{eq:zg} \\
\chi_\alpha^{max}(m_q)&=&c_\alpha m_q^{-z_\alpha}. \label{eq:zm}
\een
where the index $\alpha$ labels various susceptibilities defined in
(1--6).
Our results for exponents obtained by a single-power fit
are summarized in Table~\ref{tab:exponents}.  For $L=8$  our values
are consistent with those of Ref.~\cite{karschlaermann} carried out
for $L=8$ and $0.02\leq m_q\leq 0.075$.

Th exponent $z_g$ determined from the critical coupling is roughly
consistent with the O(2) or O(4) value theoretically expected.
All the other exponents exhibit a systematic
increase with $L$ and
deviate significantly from the predictions, particularly
for $z_t$ and $z_e$.

On the other hand, the hyperscaling relations $z_g+z_m=z_t+1$ and
$2z_t-z_m=z_e$, which follow from the fact that all the exponents are
determined by the basic thermal and magnetic exponents,
are well satisfied for each spatial size $L$.

We have also calculated the scaling function $F_m(x)=m_q^{z_m}\chi_m(g^2,m_q)$
with $x=(g_c^{-2}(m_q)-g_c^{-2}(0))\cdot m_q^{-z_g}$.  We find reasonable
scaling if measured values of exponents are employed,
while results are much worse with the use of the O(4) exponents.

We may summarize that our susceptibility data are consistent with
a second-order transition, but that the exponents apparently take
values different from those theoretically expected, at least in the range
of quark mass $m_q\geq 0.01$.

\section{Concluding remarks}

This investigation has raised several issues which were not apparent in the
previous studies\cite{founf2,columbianf2,karschlaermann}.
While we feel a first-order transition being unlikely, a finite-size
analysis is needed at a larger spatial volume to  firmly establish
the absence of a first-order transition at $m_q=0.01$.
The discrepancy of measured values of exponents from
the theoretical expectations, especially from those of O(2), has to be
clarified to confirm the second-order nature of the transition.

\vspace*{2mm}
This work is supported by the Supercomputer Project (No. 97-15)
of High Energy Accelerator Research Organization (KEK),
and also in part by the Grants-in-Aid of
the Ministry of Education (Nos. 08640349, 08640350, 08640404,
09246206, 09304029, 09740226).

\end{document}